\documentclass[global,twocolumn]{svjour}

\usepackage{latexsym}
\usepackage{graphics}
\usepackage{textcomp}
\usepackage[dvips]{graphicx}
\usepackage[dvips]{epsfig}

\journalname{Appl. Phys. B}

\begin{document}

\title{Bose-Einstein Condensation in a CO$_2$-laser Optical Dipole Trap}
\author{Giovanni Cennini, Gunnar Ritt, Carsten Geckeler, and Martin Weitz}

\institute{Physikalisches Institut der Universit\"at T\"ubingen,
Auf der Morgenstelle 14, 72076 T\"ubingen, Germany }

\date{Received: date / Revised version: date}

\maketitle
\begin{abstract}
We report on the achieving of Bose-Einstein condensation of a
dilute atomic gas based on trapping atoms in tightly confining
CO$_2$-laser dipole potentials. Quantum degeneracy of rubidium
atoms is reached by direct evaporative cooling in both crossed and
single beam trapping geometries. At the heart of these all-optical
condensation experiments is the ability to obtain high initial
atomic densities in quasistatic dipole traps by laser cooling
techniques. Finally, we demonstrate the formation of a condensate
in a field insensitive $m_{F}=0$ spin projection only. This
suppresses fluctuations of the chemical potential from stray
magnetic fields.
\end{abstract}
\section{Introduction}
\label{intro} Since the first observation of Bose-Einstein
condensation (BEC) in a dilute atomic gas, we have been witnessing
dramatic progress in both theoretical and experimental studies of
degenenerate quantum gases \cite{cornell,ketterle}. A variety of
static and dynamic properties of the condensed, trapped atoms have
been exploited \cite{stringari}. Interference studies with
Bose-condensates have experimentally demonstrated the coherence
properties expected for such a `giant matter wave'
\cite{ketterle}. More recently, the Mott-insulator phase
transition has been observed by loading a magnetically trapped
condensate into a steep optical lattice \cite{immanuel}.

The standard approach to produce a Bose-Einstein condensate
involves a laser cooling of the atomic sample to a phase space
density of $10^{-6}$, after which atoms in weak-field seeking
states are transferred into a magnetic trap
\cite{cornell,ketterle}. Further cooling proceeds by a
radiofrequency field selectively transferring the highest
energetic atoms into untrapped states, which evaporatively cools
the atoms to quantum degeneracy. A drawback of magnetic traps is
that they only confine weak field seeking spins states, which e.g.
prevents one from studying spinor condensates directly in these
traps \cite{ketterle}. Moreover, atoms in field independent
$m_{F}=0$ spin projections, as are of large interest for many
precision experiments \cite{metrology}, cannot be confined in
magnetic traps.

Far detuned optical dipole traps can confine atoms in arbitrary
spin states. Dipole traps furthermore allow for very variable
trapping geometries, as e.g. periodic lattice geometries.
Experiments aiming at an achieving of quantum degeneracy in
optical dipole traps have been first carried out by the Stanford
group \cite{chuevap}. In far detuned dipole traps, a variety of
refined optical and evaporative cooling techniques have been
implemented, which has allowed for atomic phase space densities
very close to quantum degeneracy \cite{longlist}. Friebel et al.
found that in quasistatic dipole traps, as can be realized with a
CO$_2$-laser, polarization gradient cooling alone can accumulated
atoms to a phase space density three orders of magnitude above
that of a conventional MOT \cite{susanne2}. Chapman and coworkers
have reported on the observation of Bose-Einstein condensation by
direct evaporative cooling in a crossed CO$_2$-laser dipole trap
\cite{chapman}. All-optical cooling techniques have also enabled
the first observations of quantum degeneracy of $^{6}\mbox{Li}$,
cesium and ytterbium atoms \cite{thomas,grimm,Takasu}.

We here describe experiments creating rubidium Bose-Einstein
condensates by direct evaporative cooling in both a crossed and a
single beam quasistatic dipole trapping geometry \cite{DPG03}. Due
to the choice of a tight trapping laser focus, stable evaporation
to a Bose-Einstein condensate is possible in both geometries even
with moderate initial number of atoms. Especially the single
dipole trapping geometry allows for a considerable experimental
simplification of methods to produce Bose-condensed atoms. Note
that previous experiments realizing quantum degeneracy in dipole
traps have either required a more alignment sensitive crossed beam
geometry or Feshbach resonances to enhance the collisional rate
\cite{chapman,thomas,grimm,Takasu}. By applying a magnetic field
gradient to the single beam configuration, we realized a trap
stable for atoms in a field insensitive $m_{F}=0$ spin projection
only. In a recent work, this has enabled us to realize an
all-optical atom laser \cite{cennini}. The chemical potential of
the generated condensate is first order insensitive to
fluctuations from stray magnetic fields.

\section{Quasistatic Optical Dipole Taps}
\label{sec:1} Optical dipole traps for atoms are based on the
force which arises from the coherent interaction between the
induced atomic dipole moment and the laser electric field
\cite{chu,cohen}. When the laser frequency is tuned to the red
side of an atomic resonance, the dipole force pulls atoms towards
the intensity maxima of the field. For optical fields with
frequency far below all electric dipole resonances of ground state
atoms (i.e. `quasistatic' fields), the trapping potential is given
by\begin{center}\begin{equation}\label{1}
    U=-\frac{1}{2}\alpha_{\mathrm{s}}\left|\vec{E}\right|^2
\end{equation}\end{center}
where $\alpha_{\mathrm{s}}$ denotes the static atomic polarizability of the
electronic ground state and $\vec{E}$ is the electric field of the
optical wave.
For the ground state and the
first excited state of the rubidium atom, the scalar static
polarizability is
 $\alpha_{5S}=5.3\times10^{-39}\,\mathrm{m^2 C/V}$ and
 $\alpha_{5P}=1.3\times10^{-38}\,\mathrm{m^2 C/V}$, respectively \cite{polarizability}.
Since the frequency of the CO$_2$-laser radiation ($\lambda$ is
near $10.6\,\mbox{\textmu m}$) used in our experiment is
approximately one order of magnitude below that of a typical
electric dipole transition of ground state alkaline atoms, the
photon scattering rate is very small. In particular, the Rayleigh
photon scattering rate in this quasistatic regime is given by the
expression \begin{center}\begin{equation}\label{2}
    \Gamma_{\mathrm{s}}=\frac{16r_0^2P}{3\hbar{w^2_0}}\left(\frac{m_{\mathrm{e}}\alpha_{\mathrm{s}}}{e^2}\right)\omega^3
\end{equation}\end{center}
where $w_0$ is the beam waist of a laser beam with power P, $r_0$
the classical electron radius, $m_{\mathrm{e}}$ the electron mass
and $\omega$ denotes the optical frequency of the laser light. In
our experiments, the Rayleigh scattering rate of trapped rubidium
atoms is of the order of $1/600\,\mathrm{s}$. This results in a
small coupling of the atoms to the environment, so that the effect
of decoherence can be kept small. Let us note that for alkali
atoms trapped in an extremely far-detuned laser field, the
detuning from resonance is much larger than both the hyperfine and
fine structure splitting.  This condition is met in our
quasistatic trapping field. The trapping potential for
ground-state alkali atoms here corresponds to that of a $J=0$ to
$J'=1$ transition, and the ac Stark shift for all ground-state
sublevels is almost identical. This state-independent confinement
of extremely far-detuned optical dipole traps for many experiments
represent an advantage over magnetic traps. It is however clear
that evaporative cooling cannot be achieved by selectively
transferring atoms into untrapped states, as usually done in
magnetic traps. Such a cooling in optical traps can be achieved by
lowering the intensity of the trapping beams in a controlled way,
which decreases the depth of trapping potential \cite{chuevap}.

\section{Experimental setup and procedure}
\label{sec:2} 
\begin{figure}
\includegraphics[width=\linewidth]{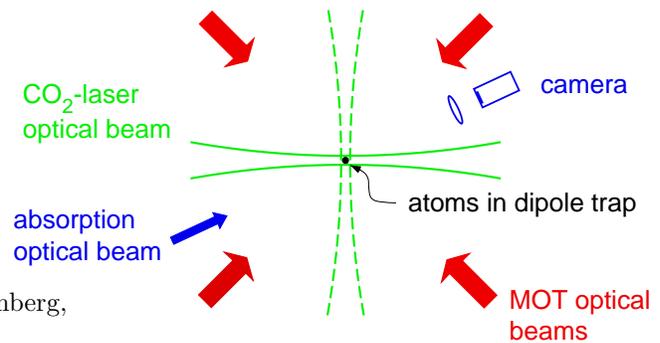}
\caption{Scheme of experimental setup}
\label{Fig:1}       
\end{figure}

A scheme of our experimental setup used for direct
Bose-Einstein condensation of rubidium atoms in a dipole trap is
shown in Fig.~\ref{Fig:1}. A commercial, rf-excited single mode
CO$_2$-laser generates up to $50\,\mathrm{W}$ mid-infrared
radiation near $10.6\,\mbox{\textmu m}$. The light passes two
acoustooptic modulators (AOMs). The first one allows us to
regulate the laser beam intensity via controlling of the rf-drive
power. The light diffracted in first order from this modulator
passes a second AOM, whose function is that of a beam divider for
the crossed beam experiments. Note that the zero's and first order
diffracted beams differ in frequency, which eliminates unwanted
standing-wave effects. The mid-infrared beams enter a vacuum
chamber through ZnSe windows. For each beam, an adjustable,
spherically corrected ZnSe lens ($f=38.1\,\mathrm{mm}$) placed
inside the vacuum chamber focuses the laser beams to a minimum
waist size of $20\,\mbox{\textmu m}$. The beam focus can be varied
with a telescope located outside of the vacuum chamber. For the
crossed trap geometry, a beam waist of $35\,\mbox{\textmu m}$ is
chosen, whereas for a single running wave geometry we use a beam
size of $27\,\mbox{\textmu m}$ in the trapping region. For the
former geometry, the trapping beams are directed horizontally and
vertically respectively and cross each other with a 90 degree
angle. For the single beam trapping geometry, the second AOM is
omitted and the transmitted beam travels through the vacuum can in
horizontal direction.

Atoms of the isotope $^{87}\mbox{Rb}$ are initially collected and
pre-cooled in a magneto-optical trap (MOT), which is loaded from
the thermal gas emitted by heated rubidium dispensers. The
measured background gas pressure inside the vacuum chamber is
$1\times10^{-10}\,\mathrm{mbar}$. The near-resonant radiation for
cooling and trapping atoms in the MOT is generated by grating
stabilized diode lasers. A cooling laser operates with variable
red detuning from the closed $5S_{1/2}$, $F=2$ to $5P_{3/2}$,
$F'=3$ hyperfine component of rubidium D2-line. Its radiation is
amplified by a second injection locked free-running diode laser.
In order to repump rubidium atoms into the cooling cycle, a second
grating-stabilized diode laser locked to the $5S_{1/2}$, $F=1$ to
$5P_{3/2}$, $F'=2$ transition is employed. Both cooling and
repumping light beams pass acoustooptic modulators and are
spatially filtered. These near-resonant beams are expanded to a
beam diameter of $20\,\mathrm{mm}$, spatially overlapped and then
directed to the vacuum chamber to provide three retroreflected MOT
beams. The total optical power of the cooling light after spatial
filtering is $42\,\mathrm{mW}$, whereas that of the repumping
light is $9\,\mathrm{mW}$. We use a pair of magnetic coils
oriented in anti-Helmholtz configuration to generate a magnetic
quadrupole field with a $10\,\mathrm{G/cm}$ field gradient, whose
center is close to the intersection of the two CO$_2$-laser beams.
The coil axes is oriented at an angle of 45 degree with respect to
the vertical and horizontal CO$_2$-laser beams. This quadrupole
field is e.g. used to operate the MOT.

An experimental cycle begins by loading the magneto-optical trap.
During the course of the experiment, we have used different MOT
loading times. A typical value for e.g. the earlier crossed beam
dipole trapping experiments is $1\times 10^7$ trapped atoms in the
MOT accumulated during a 5~s loading time. During this MOT loading
phase, the cooling laser operates with a red detuning of
$18\,\mathrm{MHz}$ to the red of the cycling transition.
Subsequently, we increase the detuning of the cooling laser to a
value of $160\,\mathrm{MHz}$ and simultaneously reduce the
repumping laser intensity by a factor 100 for a $60\,\mathrm{ms}$
long time interval. In this temporal dark MOT phase
(d-MOT)\cite{DMOT}, the atoms are pumped into the lower hyperfine
state ($F=1$), which reduces hyperfine changing collisions. These
quoted parameters are optimized by maximizing the number of atoms
transferred to the dipole trap. Throughout this cycle the
CO$_2$-laser radiation is left on, allowing for rubidium atoms to
accumulate in the dipole trapping region. At the end of the d-MOT
phase, the repumping light is extinguished and after a delay of
$2\,\mathrm{ms}$ also the cooling light. Simultaneously, the
magnetic quadrupole field is switched off. By this time, the atoms
are confined by the quasistatic dipole trap alone. All
near-resonant beams are carefully extinguished by both AOMs and
mechanical shutters. The small delay between the extinguishing of
repumping and cooling light minimizes losses in the dipole trap
due to hyperfine changing collisions.

To analyze the properties of trapped atomic clouds, we employ the
technique of absorption imaging. At the end of an experiment
cycle, the trapped atoms are released by switching off the
CO$_2$-laser beams. After a variable expansion time (typically
5--$20\,\mathrm{ms}$), cold atoms are irradiated with a pulse of
light tuned resonantly to the $F=2$, $F'=3$ component of the
rubidium D2 line. For this measurement, we use a spatially
filtered laser beam with a $8\,\mathrm{mm}$ beam diameter. The
beam is then imaged onto a slow-scan CCD camera, where shadow
images of the atomic cloud are recorded. From these images, both
number of atoms and atomic temperatures are extracted. The pulse
length is $80\,\mbox{\textmu s}$ and the intensity
($100\,\mathrm{\mbox{\textmu}W/cm^2}$) is chosen to be clearly
below saturation intensity. During this detection time, repumping
light is provided by the MOT beams.

\section{Experimental Results}
\label{sec:3}

\subsection{Bose-Einstein condensation in a crossed beam geometry}
\label{subsec:3.1} 
\begin{figure*}
\begin{minipage}{\linewidth}
\includegraphics[width=0.5\linewidth]{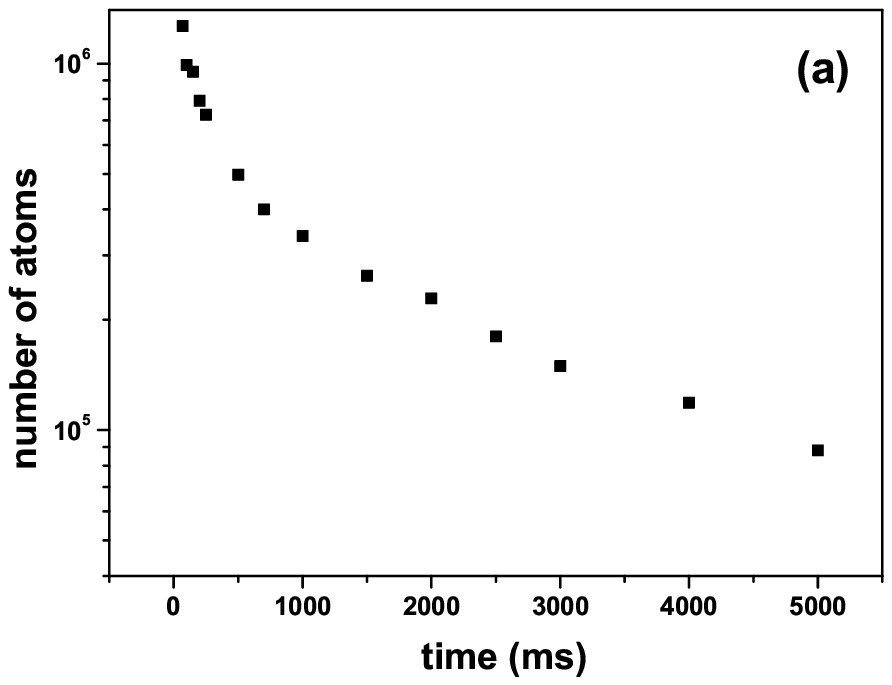}
\hfill
\includegraphics[width=0.5\linewidth]{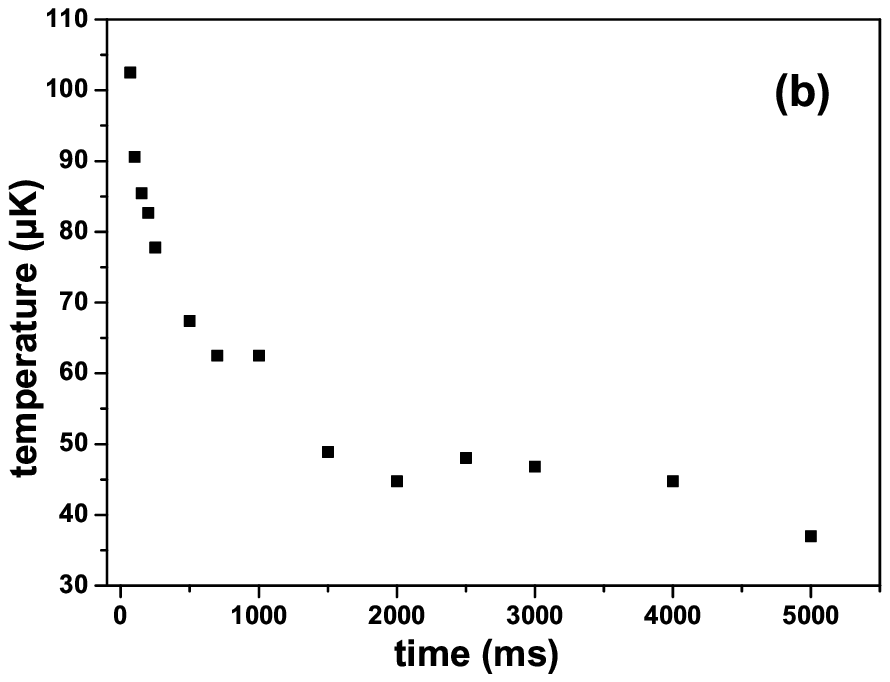}
\end{minipage}
\caption{(a): Number of trapped atoms and (b) atomic temperature
versus trapping time in a CO$_2$-laser crossed dipole trap
configuration.}
\label{Fig:2}       
\end{figure*}

\begin{figure*}
\includegraphics[width=\linewidth]{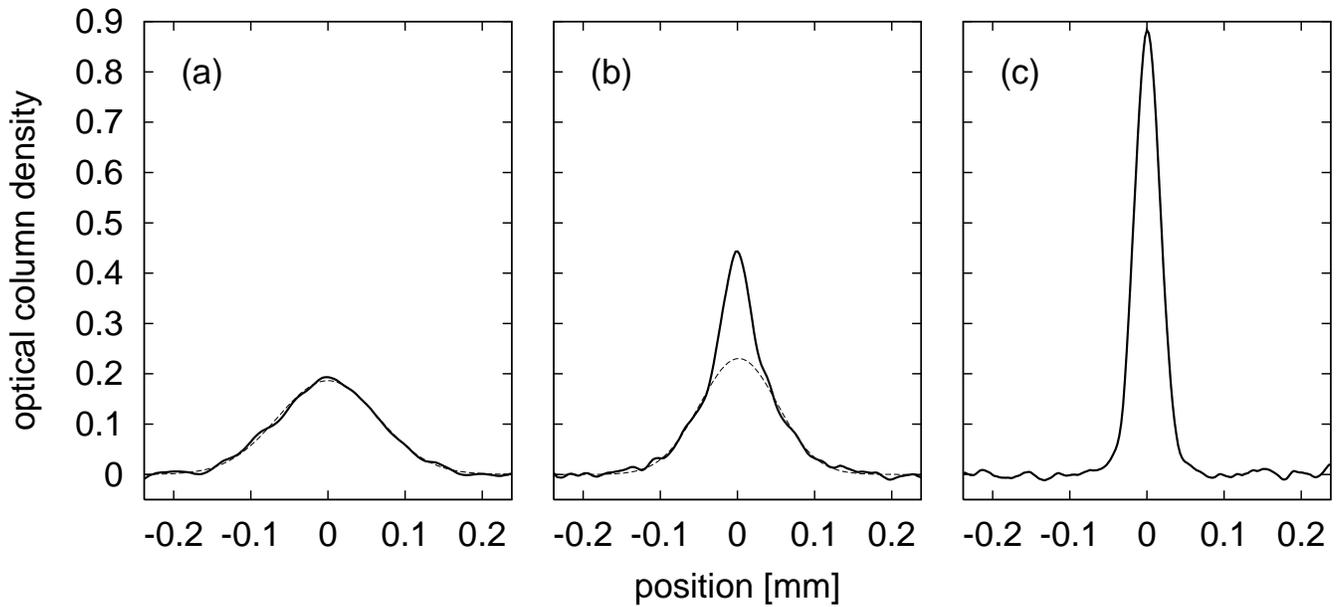}
\caption{Formation of a Bose-Einstein condensate in a crossed beam
dipole trap. The plots show density profiles for (a) a final
trapping beam power of 150 mW, yielding a thermal cloud
corresponding to a temperature of 240 nK. (b) Final ramp power of
100~mW, yielding an atomic temperature of 200~nK. This value is
near the critical temperature of 190~nK. A bimodal distribution
with the central feature corresponding to Bose-condensed atoms and
a thermal atomic distribution in the wings is visible. (c) Data
taken at for final ramp power of 75 mW. An almost pure
Bose-Einstein condensate is here produced.}
\label{Fig:3}       
\end{figure*}

In initial experiments, we studied BEC in a
crossed beam geometry, similarly as done in work by Chapman et al.
\cite{chapman}. To reach Bose-Einstein condensation by direct
evaporative cooling in our dipole traps, the initial phase space
density of the laser cooled clouds is first optimized. We have
characterized the trap lifetime, trap frequencies and the
temperature of the trapped atomic cloud. The vibrational
frequencies are measured by periodically modulating the optical
confining power with the first AOM, which allows us to study
parametrically excitation of the trapped atoms \cite{susanne1}.
Significant trap loss is observed when the modulation frequency is
close to twice a trap vibrational frequency. With typical
parameters of $12\,\mathrm{W}$ optical power in each of the
CO$_2$-laser trapping beams and a $35\,\mbox{\textmu m}$ waist
radius, we observe parametric resonance corresponding to a
vibrational frequency of $\nu=1.7\,\mathrm{kHz}$. For this crossed
beam geometry, the oscillation frequencies in the central trap
region in all three spatial directions are similar. The trap
lifetime is extracted by monitoring the number of atoms detected
after different trapping times for a constant potential depth. A
typical result of such a measurement is shown in
Fig.~\ref{Fig:2}a. The data points are taken after a minimum
dipole trapping time of 70 ms, since for shorter delay times atoms
not being transferred to the dipole trap have not yet been removed
completely from the detection region by gravity. The data can be
described with a two time constants exponential decay, which
involves a fast and a slow decay with time constants near
$100\,\mathrm{ms}$ and $12\,\mathrm{s}$ respectively. The fast
decay in the beginning of the dipole trapping phase is attributed
to a plain atomic evaporation of the highest energetic atoms. In
contrast, the slow decay is attributed to losses due to collisions
with background gas, which ultimately limit the available trapping
time. Fig.~\ref{Fig:2}b shows the dependence of the atomic
temperature on trapping time. The temperature clearly decreases
with time, which we attribute to the plain evaporation. This
process is effective only for small trapping times when the loss
rate is large. The measured atomic density after a trapping time
of $500\,\mathrm{ms}$ is
$1\times10^{13}\,\mathrm{atoms}/\mathrm{cm}^{3}$ . Together with
the measured temperature of $60\,\mbox{\textmu K}$ we derive a
phase space product $n\lambda^3_{\mathrm{dB}}\simeq1/500$, which
corresponding to a $1/1500$ phase space density if we assume equal
distribution of spin projections. These results are comparable to
previous results observed in CO$_2$-laser optical lattice and
crossed dipole geometries \cite{susanne2}. We also infer a
collisional rate of $7\,\mathrm{kHz}$ at this time, clearly
representing a favorable starting point for further, forced
evaporative cooling to lower temperatures. To induce such a forced
evaporation, the optical potential is lowered in a controlled way.
Experimentally, this is achieved by reducing the trapping laser
beam intensity via a decreasing the rf-drive power of the first
AOM. Mandatory for the success of evaporative cooling is to
maintain a relatively high atomic collision rate, so that the
remaining atoms can rethermalize to a distribution of lower
temperature. We reduce the potential depth with time according to
the formula
\begin{equation}\label{3}
    U(t)=U_0(1+t/\tau)^{-\beta} ,
\end{equation}
where $U_0$ denotes the value of the initial potential depth. It
was shown in Ref. 10 that this ramp form maintains a nearly
constant ratio $\eta=U/k_{\mathrm{B}}T$ between potential depth
and the atomic temperature. Following a 100~ms long plain
evaporation period, the mid-infrared power is reduced in a
$3.5\,\mathrm{s}$ long ramp time from initially $12\,\mathrm{W}$
to a final value of $75\,\mathrm{mW}$ in each of the trapping
beams. In these crossed dipole experiments, we find optimum
cooling when choosing the parameters of Eq. (3) to be
$\tau=0.3\,\mathrm{s}$ and $\beta=1.5$. We have experimentally
verified that the cutoff parameter $\eta$ was nearly constant
throughout the evaporation ramp. For a characterization of the
cooling, we have analyzed shadow images of the expanded atomic
cloud at the end of the evaporation process. Fig.~\ref{Fig:3}a
shows a cross-section of such an image for a final trapping power
of 150 mW, corresponding to an average trap vibrational frequency
of 350 Hz. It is clear that here the ensemble is still purely
thermal. The phase space density is further increased when we ramp
to lower power values. Fig.~\ref{Fig:3}b shows data recorded for a
final beam power of $100\,\mathrm{mW}$, where a typical bimodal
distribution corresponding to an atomic cloud near the transition
point is observed. The central feature corresponds to atoms in the
Bose-condensate, while the wings represent thermal atoms. The
critical temperature here is $T_{\mathrm{c}}=190\,\mathrm{nK}$.
For final power of $75\,\mathrm{mW}$, an almost pure condensate is
obtained as shown in Fig.~\ref{Fig:3}c. The condensate contains
about 10000 atoms distributed among the three $m_F$ states of the
electronic hyperfine ground state $F = 1$. In other measurements,
we have applied a magnetic quadrupole field during the course of
the evaporation by leading on the MOT coils. This leads to a
condensate with about 70 percent population in the $m_F=1$ and 30
percent in the $m_F=0$ spin projection. For these trap parameters,
the magnetic field gradient predominantly populates one
field-sensitive state. In earlier work in a crossed dipole trap it
was demonstrated that with different field gradients also other
spin states can be populated, although it was noted that by the
time of writing the origin of the obtained magnetization still
remained to be determined \cite{chapman_ICAP}.

\subsection{BEC in a single running wave trap}
\label{subsec:3.2} In this chapter, we describe our experiments
successfully creating a rubidium Bose-Einstein-condensate in a
single beam CO$_2$-laser dipole trap. This is the simplest
possible dipole trapping geometry. The use of single running wave
allows for a compact setup and a very reliable production of the
BEC. Previous experiments creating quantum degeneracy in dipole
traps have either required the use of a more alignment-sensitive
crossed dipole trap geometry \cite{chapman,Takasu} or Feshbach
resonances \cite{grimm} to enhance the collisional rate during the
course of the forced evaporation.

As in the case of the crossed dipole trap configuration,
evaporative cooling towards BEC requires a ramping down of the
trapping potential. Since the average of the trap vibrational
frequencies for a given power and beam waist is smaller than in
the crossed beam geometry, it is \textit{a priori} not clear that
the rethermalization rate is sufficiently high throughout the
evaporation ramp. Obviously, the final stages of evaporation here
are most critical, since the collisional rate decreases with trap
power. It is easy to show that the trap vibrational frequencies in
a single dipole trap orthogonal to the beam axis scale as
$\nu_r\propto \sqrt{P}/w^2_0$, while along the beam axes scale as
$\nu_z\propto \lambda\sqrt{P}/w^3_0$, where $\lambda$ is the laser
wavelength. Our experiments demonstrate, that by choosing a
sufficiently small beam waist, the collisional rate can be high
enough throughout the ramp, and BEC can be reached. For the single
beam geometry, the focus of the (horizontally oriented) trapping
beam is reduced to a $27\,\mbox{\textmu m}$ beam waist. In
principle, one could work with an even tighter focus to fully
compensate for the smaller compression in this geometry. In our
experiments, we have instead started with a higher number of atoms
and also used a slightly longer evaporation time. The weak
confinement along the beam axis of the single dipole trap geometry
offers the additional benefit that it allows us to remove atoms in
field sensitive spin projections with a moderate magnetic field
gradient. We note that an alternative to such a state selection is
to achieve field-insensitivity by using of atoms with a spin
singlet ground state, as e.g. atomic ytterbium. Very recently in
an impressive experiment, evaporative cooling in a crossed dipole
trap has allowed for the production of a Bose-Einstein condensate
of these atoms \cite{Takasu}.

For our single dipole trap experiments, we increase the MOT
loading time to $30\,\mathrm{s}$, during which $6\times 10^7$ are
captured. At the end of the MOT loading, we apply a temporal
dark-MOT phase as described in the above section. Subsequently,
all near-resonant optical beams are extinguished and the atoms are
trapped in the quasistatic dipole trapping potential alone.
Typically, $4\times10^{6}$ atoms are captured in the single beam
trap. We have characterized the vibrational frequencies of the
trap. At full trapping laser power ($28\,\mathrm{W}$), we measure
$\nu_r=4.8\,\mathrm{kHz}$ and $\nu_{z}=350\,\mathrm{Hz}$,
corresponding to vibrations orthogonal and collinear respectively
to the beam axes. In this single beam geometry, there is no
spherical symmetry and the longitudinal trap frequency is about 13
times smaller than the radial one. We again have first analyzed
the trapping of atoms while maintaining full CO$_2$-laser trapping
power. Following a 100\,ms long plain evaporation phase, we
measure an atomic temperature of $140\,\mbox{\textmu K}$ and an
atomic density of $n\simeq 1.2 \times
10^{13}\,\mathrm{atoms/cm^3}$. This corresponds to a product $n
\lambda^{3}_{\mathrm{dB}} \simeq 1.2 \cdot 10^{-4}$. Although this
value is below best values obtained in quasistatic crossed dipole
trap configurations or optical lattices, it is clearly above
results achieved in magnetic traps prior to forced evaporative
cooling. From the above values, we derive a high collision rate of
$6.2\,\mathrm{kHz}$. This has encouraged us to proceed with forced
evaporative cooling in this single beam dipole trapping geometry.

\begin{figure}
\includegraphics[width=\linewidth]{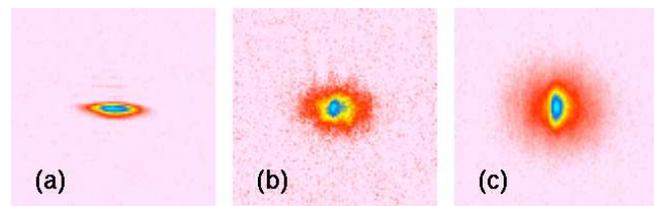}
\caption{Free expansion of a Bose-Einstein condensate generated in
a single beam dipole trap. Shown is a series of shadow images of
the atomic cloud recorded after allowing for different free
expansion times (field of view $240\,\mbox{\textmu m}\times
240\,\mbox{\textmu m}$). (a) Image taken with nearly no free
expansion, showing thus the cigar-shaped spatial distribution of
the trapped cloud. (b) and (c): Recorded after a $8\,\mathrm{ms}$
and $15\,\mathrm{ms}$ respectively of free expansion. Especially
at the latter time, the asymmetry of the cloud is inverted.}
\label{Fig:4}       
\end{figure}

\begin{figure}
\centering\includegraphics[width=0.75\linewidth]{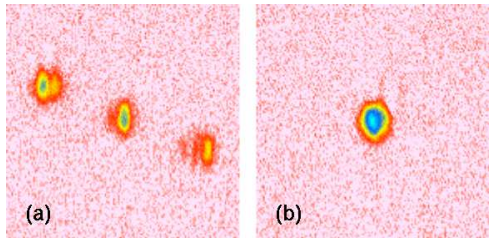}
\caption{Shadow images of Bose-Einstein condensates generated in a
single beam dipole trap after allowing 15~ms of free expansion,
see also \cite{cennini}. The field of view comprises
$380\,\mbox{\textmu m}\times 380\,\mbox{\textmu m}$. (a)
Stern-Gerlach magnetic field applied only during the free
expansion phase. The three components $m_F=\pm 1$, 0 of a spinor
condensate have separated into three spatially resolved clouds.
(b) Stern-Gerlach field activated during the evaporative cooling
phase. Here, a pure $m_F=0$ condensate is generated.}
\label{Fig:5}       
\end{figure}

To cool the atoms towards lower temperatures, we apply a
$7\,\mathrm{s}$ long forced evaporation ramp, during which the
power of the dipole trapping beam is reduced from its initial
value of $28\,\mathrm{W}$ to a final value of $200\,\mathrm{mW}$.
This ramp is applied directly after the dark MOT phase, as for
this geometry no improved cooling is observed when an initial
plain evaporation phase is added. We again use a ramp form as
described by Eq.~\ref{3}. Optimum cooling in the single dipole
trap was observed when using parameters $\tau$ and $\beta$ near
$0.45\,\mathrm{s}$ and 1.4 respectively. We have recorded
time-of-flight (TOF) shadow images of the atomic cloud for
different free expansion times. For this measurement, the MOT
quadrupole field was switched off after the d-MOT phase.
Fig.~\ref{Fig:4}a shows an image recorded directly after switching
of the trap laser, i.e. with no free expansion. The trapped cloud
is cigar shaped and strongly elongated along the weakly confining
beam axis. Figs.~\ref{Fig:4}b and \ref{Fig:4}c give time of flight
images recorded after a 8\,ms and 15\,ms respectively long free
expansion phase. In the former image the cloud is almost
symmetric, while in the latter image the symmetry axis is
inverted. We interpret this series of images as a clear evidence
for Bose-Einstein condensation in the single beam optical trap. In
this asymmetric trap geometry, the cloud expands faster along the
radial than along the axial direction due to the anisotropic
release of mean field energy. The formed condensate has spinor
nature. To analyze the distribution of Zeeman components, we apply
a Stern-Gerlach magnetic field gradient during the free expansion
phase, which spatially separates the different Zeeman components.
Fig.~\ref{Fig:5}a shows a typical measurement, where a separation
into clouds with different spin projections is clearly visible. We
produce condensates with typically 12000 atoms distributed among
the $m_{F} = -1, 0, 1$ Zeeman components of the $F = 1$ ground
state. As stated above, one of the main advantages of optical
dipole traps is their state independent trapping, which allows for
the formation and confinement of spinor condensates.

In subsequent experiments, we have condensed atoms in the $m_F=0$
spin projection only. For this measurement, the magnetic
quadrupole field generated by the MOT coils was left on throughout
the experimental cycle, i.e. also during both dipole trapping and
detection phases. For our experimental parameters, stable trapping
of atoms during the final phase of the evaporative cooling is then
only possible for the field insensitive ($m_F=0$) spin projection.
Atoms in spin projections $m_F=\pm 1$ are removed during the
course of evaporative cooling by the field gradient (note that the
dipole trap does in general not exactly overlap with the center of
the quadrupole field). Fig.~\ref{Fig:5}b shows a typical obtained
time-of-flight shadow image of the condensed cloud. The
measurement shows that only the $m_F=0$ Zeeman state is populated.
We typically obtain 7000 condensed atoms in this first order
field-insensitive state. Note that this number is larger than the
obtained $m_F=0$ fraction of the above discussed spinor
condensate, which we attribute to sympathetic cooling with atoms
in $m_F=\pm 1$ spin projections during condensate formation. The
critical temperature of the $m_F=0$ condensate is
 $T_{\mathrm{c}}\simeq 220\,\mathrm{nK}$, and the
peak-density is $1.2\times 10^{14}/\mathrm{cm}^3$. For both the
spinor and the $m_F=0$ condensate we measure condensate lifetimes
near $5\,\mathrm{s}$. This value is shorter than the inverse loss
rate due to background gas collisions, and we attribute this
lifetime to be mainly limited by three-body losses. Within our
experimental uncertainty, we do not observe any difference between
the lifetimes of a spinor and the $m_F=0$ condensate. Further, we
do not observe transfer of atoms from $m_F=0$ state into
field-sensitive states. To our belief, spin changing collisions
here are suppressed by the quadratic Zeeman shift in the
inhomogeneous magnetic field \cite{greene,grimmcomm}.

\section{Conclusions and Outlook}
\label{sec:4}

We have investigated Bose-Einstein condensation of rubidium atoms
by direct evaporative cooling in tightly confining, quasistatic
dipole traps. The quantum degenerate regime has been reached both
in a crossed beam and a single beam CO$_2$-laser dipole trapping
geometry. The former configuration allows cooling to quantum
degeneracy with a somewhat larger trap focus. On the other hand,
the newly developed single beam approach towards Bose-Einstein
condensation provides a simpler setup and is less alignment
sensitive. The scheme provides a very stable route towards the
achieving of quantum degeneracy. Finally, we demonstrated that in
the single beam setup a moderate magnetic field gradient yields a
trapping configuration, in which stable confinement is possible
only for atoms in a $m_F=0$ spin projection. The chemical
potential of the produced Bose-Einstein condensate is first order
insensitive to stray magnetic fields.

For the future, we believe that the demonstrated single beam
method for producing Bose-Einstein condensates due to its
technical simplicity opens up wide applications in the physics
with quantum gases. Further research could investigate an
extension to the Bose-Einstein condensation of other atomic
species, as e.g. alkaline-earth atoms.

Further, we anticipate the described method to produce condensates
in a field-insensitive spin state can have impact on the
application of quantum degenerate atoms to precision measurements,
as e.g. atomic clocks or atom interferometry experiments. One can
furthermore envision experiments with two-component quantum
degenerate gases driven by exciting the microwave clock transition
between the hyperfine ground states $F = 1$, $m_F = 0$ and $F =
2$, $m_F = 0$. It clearly remains important to study the stability
of different spin projections against spin changing collisions in
more detail.

Quasistatic optical dipole traps are of large interest for quantum
computation. In a CO$_2$-laser lattice geometry the lattice
spacing is so large that individual sites can be optically
resolved, which allows for an individual addressing of qubits
\cite{rainer}. By inducing a Mott insulator phase transition,
unity occupation of atoms in lattice sites can be achieved. An
entanglement of atoms in the lattice is e.g. possible by cold,
controlled collisions \cite{jaksch}.

This work has been supported in parts by the Deutsche
Forschungsgemeinschaft, the Landesstiftung Baden W\"urttemberg,
and the European Community.
%

%

%
%

\clearpage
\end{document}